\pdfoutput=1

\PassOptionsToPackage{sort}{natbib}
\documentclass[preprint,5p]{elsarticle}

\usepackage{graphics} 
\usepackage{graphicx}
\usepackage{amsmath} 
\usepackage{amssymb}  
\usepackage{booktabs}
\usepackage{caption}
\usepackage{color}
\usepackage{subcaption}
\usepackage{changepage}
\usepackage{verbatim}
\setcitestyle{authoryear,open={(},close={)}}

\newcommand\norm[1]{\left\lVert#1\right\rVert}

\newcommand{\specialcell}[2][c]{%
  \begin{tabular}[#1]{@{}c@{}}#2\end{tabular}}

\begin{document}

\begin{abstract}

We present a method to discover differences between populations with respect to the \emph{spatial coherence} of their oriented white matter microstructure in arbitrarily shaped white matter regions. This method is applied to diffusion MRI scans of a subset of the Human Connectome Project dataset: 57 pairs of monozygotic and 52 pairs of dizygotic twins. After controlling for morphological similarity between twins, we identify 3.7\% of all white matter as being associated with genetic similarity (35.1k voxels, $p < 10^{-4}$, false discovery rate 1.5\%), 75\% of which spatially clusters into twenty-two contiguous white matter regions. Furthermore, we show that the orientation similarity within these regions generalizes to a subset of 47 pairs of non-twin siblings, and show that these siblings are on average as similar as dizygotic twins. The regions are located in deep white matter including the superior longitudinal fasciculus, the optic radiations, the middle cerebellar peduncle, the corticospinal tract, and within the anterior temporal lobe, as well as the cerebellum, brain stem, and amygdalae.

These results extend previous work using undirected fractional anisotrophy for measuring putative heritable influences in white matter. Our multidirectional extension better accounts for crossing fiber connections within voxels. This bottom up approach has at its basis a novel measurement of coherence within neighboring voxel dyads between subjects, and avoids some of the fundamental ambiguities encountered with tractographic approaches to white matter analysis that estimate global connectivity. 
\end{abstract}


\begin{frontmatter}
\title{
Spatial Coherence of Oriented White Matter Microstructure: \\Applications to White Matter Regions Associated with Genetic Similarity
}

\author[cs]{Haraldur T. Hallgr\'imsson\corref{cor}}
\ead{hth@cs.ucsb.edu}

\author[psych]{Matthew Cieslak}

\author[cs]{Luca Foschini\fnref{partly}}

\author[psych]{Scott Grafton}

\author[cs]{Ambuj K. Singh}

\cortext[cor]{Corresponding author at: Department of Computer Science, University of California, Santa Barbara, CA 93106, USA}
\fntext[partly]{Present address and affiliation: Evidation Health, Inc., 1036 Santa Barbara Street, Santa Barbara, CA 93101, USA}
\address[cs]{Department of Computer Science, University of California, Santa Barbara, CA 93106, USA}
\address[psych]{Department of Psychological and Brain Sciences, University of California, Santa Barbara, CA 93106, USA}

\end{frontmatter}

\section{Introduction}

Structural connectomics of the human brain is increasingly recognized as an essential complement to functional imaging. Imaging the physical connectivity in the brain is primarily based on diffusion-weighted MRI (dMRI). While this imaging continues to improve in angular resolution of diffusion signals, there remain significant challenges in image reconstruction, representation of diffusion features, and statistical analysis of white matter structures across populations of subjects.

There is an abundance of methods for analyzing dMRI, many of which show promise in diagnosing brain abnormalities such as strokes~\citep{yeh2013diffusion} and discovering correlates of many cognitive processes including metacognition~\citep{baird2015regional}.  Current approaches generally fall into one of two categories: \emph{Brain Graph} methods~\citep{bullmore2009complex} use dMRI to estimate ``connection strength'' between pairs of cortical regions while \emph{Scalar-based} methods calculate a single value at each voxel that is interpreted as reflecting ``white matter integrity''~\citep{jones2013white}. 

Brain Graphs succinctly represent long-range connectivity between non-overlapping parcels of gray matter. The analyst chooses a gray matter parcellation, then uses a tractography algorithm to trace paths across white matter voxels. There are many approaches to tractography, but they all utilize diffusion orientation information to grow streamlines through space.  Tractography results therefore depend on the accuracy of the voxel-wise estimates of white matter orientation, which can be complicated in structures such as crossing fibers~\citep{jbabdi2011tractography} with~\cite{maier2016tractography} suggesting these tractography methods are readily dominated by false positive streamlines. Brain Graphs represent cortical regions as nodes and use a property of streamlines (such as their count) to weight edges, resulting in a connectivity matrix. These connectivity matrices are the basis for numerous network-based approaches~\citep{bullmore2011brain} that have shown promise in understanding the development of large-scale brain connectivity~\citep{hagmann2010white} and processes such as aging, disease, and cognition~\citep{deary2006white,poudel2014white}.

\begin{figure}
\centering
\includegraphics[width=0.3\textwidth]{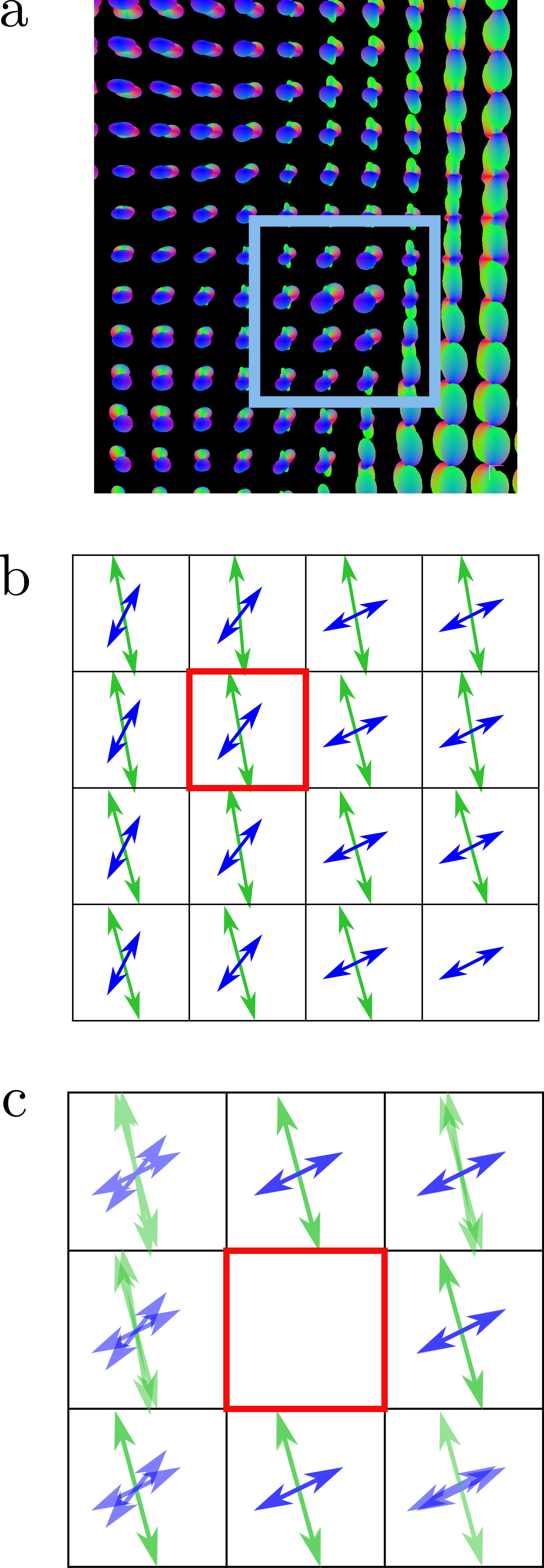}
\caption{Measuring coherence across voxels within a single subject. (a) A two-dimensional slice of the measured Orientation Distribution Function  (ODF) from a single subject measuring the Brownian motion of water that is constrained by oriented white matter microstructure. (b) The multidirectional anisotrophy (MDA) values extracted from the local peaks of the ODFs (from pink box in \emph{a}). (c) Measuring the coherence of neighboring voxels with respect to their ODFs by overlaying the extracted MDA vectors from the center voxel (highlighted in red in \emph{b}) onto all spatially adjacent white matter voxels in this 2D slice.}
\label{fig:cartoon}
\end{figure}

Scalar-based methods typically reduce the 6-dimensional dMRI data, a 3D oriented diffusion field measured in a 3D space, into a 3D volume. These scalar-valued volumes can easily be spatially normalized and statistically compared across individuals. The most common example of this is the analysis of Fractional Anisotropy (FA) derived from diffusion tensor imaging (DTI).  FA is a function of the eigenvalues of a fitted diffusion tensor, with higher values reflecting a large degree of diffusion along a single orientation while lower values can reflect white matter damage \citep{papadakis1999study,wieshmann1999reduced,filippi2001diffusion,kanaan2005diffusion,werring2000diffusion,witwer2002diffusion} or the presence of fiber populations projecting in multiple orientations \citep{jones2013white,Volz2017}. The inability of tensors to represent multiple directions has been addressed by methods that use higher angular-resolution dMRI to calculate an  orientation distribution function (ODF) in each voxel where multiple fiber populations appear as ``lobes''~\citep{wedeen2005mapping}, as seen in Fig.~\ref{fig:cartoon}a. Although ODFs can represent multiple fiber orientations, popular ODF-based scalars such as generalized fractional anisotropy (GFA)~\citep{tuch2004q} and multidimensional anisotropy (MDA)~\citep{tan2015multi} are still heavily reduced in voxels with fiber crossings~\citep{Volz2017}.  A major benefit to scalar-based techniques is that 3D interpolation can be performed accurately during spatial normalization, whereas interpolating 6D ODFs has been shown to systemically affect tractography~\citep{greene2017effect}; this normalization is important to be able to compare the same spatial regions of the brain between subjects that, in general, have different brain shapes and sizes.

Although the resampling of entire ODFs after applying a spatially-normalizing displacement field can produce undesirable results~\citep{christiaens2012effect}, 3D vector fields are generally well-behaved when spatially warped. We can take advantage of this by extracting directional maxima from each ODF and treating them as vectors.  One vector is produced from each lobe of each ODF and warped to a group template where they can be compared across subjects. We calculate a similarity measure between each voxel and its neighbors instead of performing tractography on this spatially-normalized vector field. Where tractography seeks to determine whether axons \emph{project into} a neighboring voxel, similarity scores reflect whether two voxels \emph{are part of the same white matter structure}; this can be considered a generalization of tractography, capturing both the projections and cross-sections of a single white matter structure.

\begin{figure}
\centering
\includegraphics[width=0.9\linewidth]{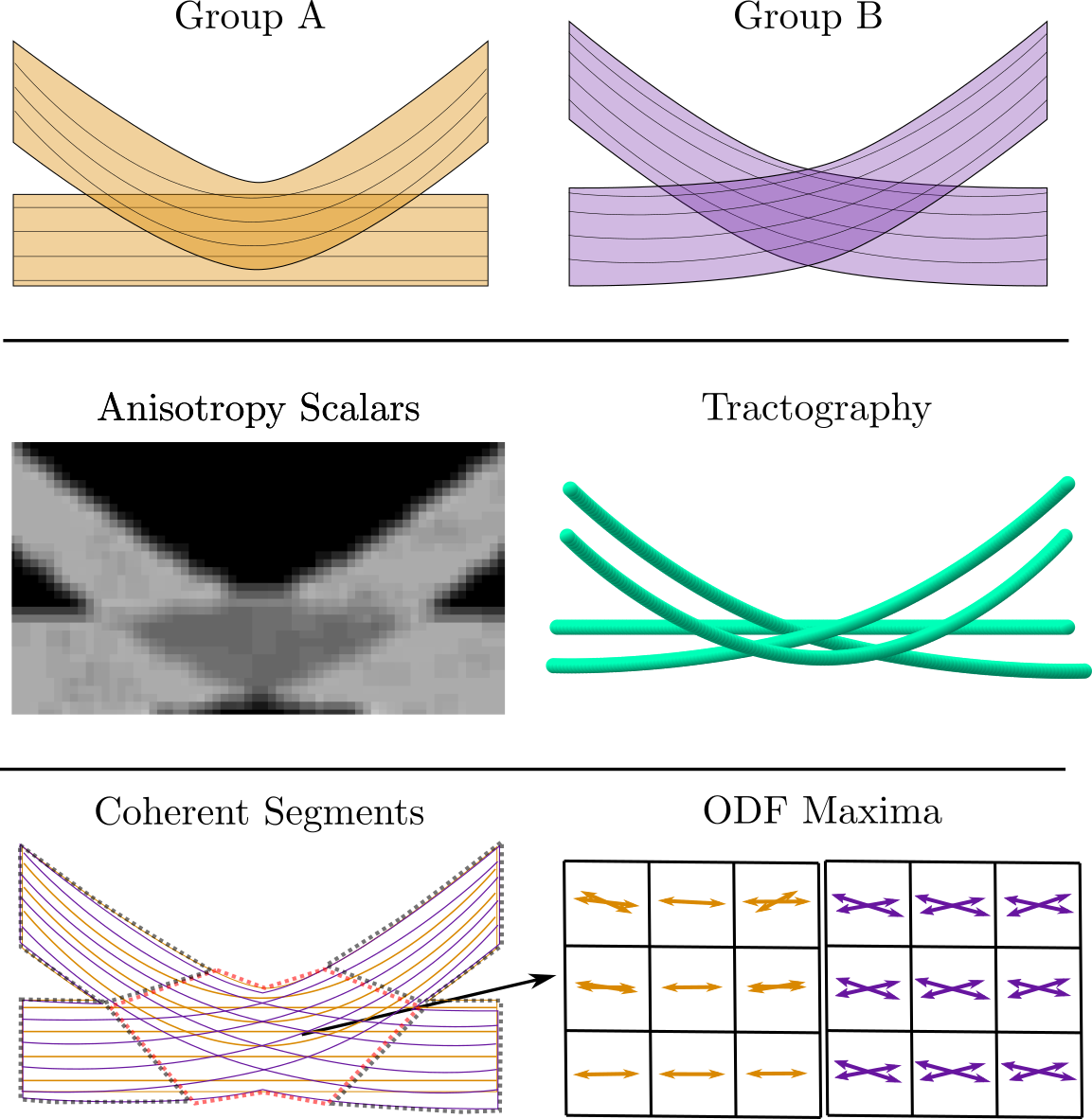}
\caption{Two fascicles from two hypothetical groups of individuals (top row). These fascicles would generate very similar anisotropy images and tractograms (middle row). Coherent regions can be identified that agree across groups (bottom left, gray outlined) and that are dissimilar across groups (red outline in center). A sample of the MDA vectors of each population from the dissimilar region is shown on the bottom right.}
\label{fig:motivation}
\end{figure}

Fig.~\ref{fig:motivation} shows how this approach compares to other current methods. Consider two fascicles in the brain that have been spatially normalized to overlap in space (top row). Two groups have different projections even though scalar based measures and tractography (middle row) would look identical. The bottom row shows the fascicles from both groups superimposed on one another.  Distance measures between neighboring voxels would reveal four areas that are \emph{coherent} both between and across groups. In contrast, the vectors in the center crossing region are coherent within each group but differ across groups. The output of this pipeline is a set of regions like the red outlined area of crossings in Fig.~\ref{fig:motivation}, where directed ODF maxima are similar within groups but differ across groups.

\begin{figure}
\centering
\includegraphics[width=0.4\textwidth]{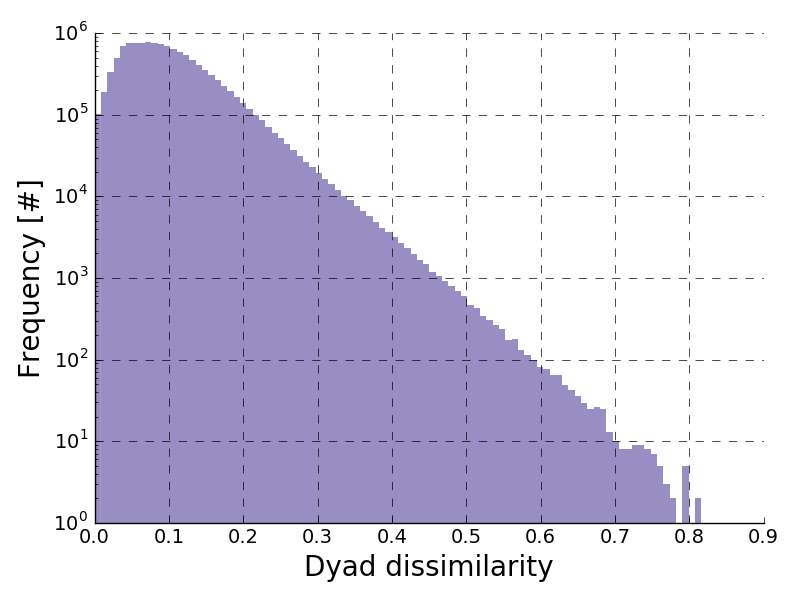}
\caption{Distribution of dissimilarity, or incoherence, between all adjacent white matter voxels within a single subject. Incoherence is mostly small but a long tail of dissimilar neighboring voxels exists.}
\label{fig:subj_coherence}
\end{figure}

Directional ODF maxima tend to vary smoothly in space albeit with large discontinuities around anatomical features, as seen in Fig.~\ref{fig:cartoon}. We can measure the similarity of neighboring voxels by defining a distance between two ODFs that takes into account both magnitude and direction of each peak. Fig.~\ref{fig:subj_coherence} shows an example of \emph{incoherence}, or dissimilarity, between ODFs from all dyads of neighboring white matter voxels within a single subject. Most dyads exhibit very low dissimilarity, with a long tail of voxel dyads with large dissimilarities. \emph{Dyadic distances form the basis for the method proposed here. These distances are used to build a lattice network, which expands the comparison from neighboring voxels to large white matter regions. Region-based distances are then used to compare between groups.}

To demonstrate the validity and usefulness of the dyad approach, we consider the problem of finding spatially contiguous regions of white matter that are associated with a population of interest as compared to some control population. This problem mirrors the approach taken to subdivide the gray matter of the brain into functional regions~\citep{glasser2016multi,zhu2011discovering}. We develop a non-parametric method for discovering arbitrarily shaped white matter regions that are significantly more similar with the population of interest, not on the basis of their connectivity to gray matter regions but instead on a group-wise local consistency in oriented white matter microstructure. This is accomplished by discovering spatially contiguous white matter voxels that are significantly more coherent within the population than would be expected from a matched control group. (Alternatively, especially in the context of neurological disorders and/or injuries, one could additionally search for regions that are \emph{less} coherent in the population of interest.) In contrast to previous studies~\citep{yeh2016connectometry}, we measure coherence simultaneously across both subjects \emph{and} neighboring voxels.

We apply this method to diffusion scans from the Human Connectome Project on a population of monozygotic (MZ) and dizygotic (DZ) twins to discover white matter regions that are associated with genetic similarity and/or a common upbringing. We hypothesize that in this situation, there should be significantly more coherence in the MZ twins than DZ twins, and DZ twins than unrelated individuals. The discovered regions are more similar within MZ and DZ twins than as compared to a control population of strangers. We also test the robustness of the discovered areas by generalizing them to a previously unseen population of non-twin siblings that display as much similarity in the white matter regions as the DZ twins.

Previous work has identified genetic influences of various quantitative measures of the brain, including total brain volume~\citep{posthuma2000multivariate}, the volume of gray and white matter regions~\citep{peper2007genetic,de2017same}, peaks of fiber orientation functions and tracts derived therefrom~\citep{shen2014investigating}, brain asymmetries~\citep{haberling2013asymmetries,jahanshad2010genetic}, as well as other aspects of white matter~\citep{yap2011development,thompson2001genetic,sadeghi2017twin}. A simpler representation of the ODFs, FA, which measures how anisotropic a ODF is, has previously been shown to measure putative heritable influences~\citep{chiang2011genetics}. Our work extends that work by considering a richer representation of the ODFs. It is important to note that in our work we are interested in similarity associated with oriented white matter microstructure and not gross anatomical morphology, which we control for by excluding those white matter voxels whose log Jacobian determinant obtained during spatial normalization are more similar in MZ and DZ twins as opposed to strangers, see~\ref{sec:appendix_jacobians}.

If the population of interest does not correspond to any significant or strong signal of similarity within the oriented white matter microstructure, our method would identify only sparse and spatially distributed portions of white matter with an associated high false discovery rate. We apply our method on a population of MZ and DZ twins as we expect, and demonstrate, this population of interest as having a very strong signal to validate our method. That the results we present cluster spatially to a very high extent, are associated with a low false discovery rate and large effect sizes, and generalize to a previously unseen portion of the population serves as a strong validation of the method.

\section{Methods}
\label{sec:methods}

\subsection{Imaging data and preprocessing}

The preprocessing pipeline used for this study was identical to that used in~\citep{Volz2017}, but is reported here as well for completeness. These data were collected as part of the Washington University-Minnesota Consortium Human Connectome Project~\citep{VanEssen2012,van2013wu,feinberg2010multiplexed}. Participants were recruited from Washington University (St. Louis, MO) and surrounding area. All participants gave informed consent. The data is derived from 630 participants (358 female, 272 male).  

The structural and diffusion data were collected on 3T Connectome Skyra system (Siemens, Erlangen, Germany). The diffusion volumes were collected with a spatial resolution of 1.25x1.25x1.25 mm$^3$, using three shells at $b = 1000$, $2000$, and $3000$ s/mm$^2$ with 90 diffusion directions per shell and 10 additional b0s per shell. Spatial distortion and eddy currents were corrected using information from acquisitions in opposite phase-encoding directions, as well as head motion~\citep{glasser2013minimal}. The high-resolution structural T1 weighted and T2 weighted volumes were acquired on the same scanner at 0.7mm isotropic resolution. Minimally preprocessed images were reconstructed in DSI Studio (http://dsi-studio.labsolver.org) using Generalized Q-Sampling Imaging~\citep{Yeh2010}.

Skull stripped, aligned, and distortion corrected T1w and T2w volumes~\citep{glasser2013minimal} were rigidly registered to the subject’s GFA volume. The symmetric group wise normalization (SyGN) method implemented in Advanced Normalization Tools (ANTs, http://\allowbreak stnava.\allowbreak github.\allowbreak io/\allowbreak ANTs/) was used to construct a custom multimodal brain template using the data of 38 HCP subjects~\citep{avants2010optimal} that included proportions of racial, gender, and handedness that chosen through stratified random sampling according to these features. Of those 38, seven are monozygotic twins and nine are dizygotic twins that are a part of the population of interest for this study, and a further four are part of the non-twin siblings set. Each subject's GFA, T1w, and T2w volumes were used during template creation with weighting factors of  0.5 (GFA) $\times$ 1 (T1w) $\times$ 1 (T2w). Templates were created after 5 iterations. Templates from the 4th and 5th iterations of multi-modal template construction were inspected to check that the templates had stabilized.
All individual datasets were ultimately normalized to this template using all 3 modalities and symmetric diffeomorphic normalization (SyN) as implemented in ANTs~\citep{avants2008symmetric}.

\subsection{Extracting MDA Vectors}
\label{sec:mda}

Each ODF $\Psi(\theta)$ was calculated with GQI on a set of 642 approximately-evenly spaced directions $\theta \in \Theta$ on a tesselated icosahedron. ODF magnitudes were rescaled so that the sum of each ODF is $\sum_{\theta \in \Theta} \Psi(\theta) = 1$. We then calculated the multi-directional anisotropy (MDA) value for each direction $\theta$ as 

\begin{equation}
\text{MDA}(\theta) = \frac{1-\mu}{\sqrt{1+2\mu^2}}
\end{equation}
where
\begin{equation}
\mu = \left( \frac{\Psi(\theta)}{\Psi(\theta_{min})} \right)^{2/3}
\end{equation}
and $\theta_{min}$ is the direction with the smallest ODF magnitude.  

MDA values were calculated for the four largest local maxima in every ODF, resulting in values denoted MDA0, MDA1, MDA2, and MDA3 which are ordered by decreasing size. The four corresponding directions $\theta_0,\theta_1,\theta_2,\theta_3$ were also extracted and saved as 3D vector fields for each of $\theta_0,\dots\theta_3$. In a separate study of the same data we found that ODF peaks become very noisy after the 4th direction~\citep{Volz2017}. Vector fields corresponding to the local maxima were warped into the group template using ANTs. 3D volumes containing MDA0-3 were also warped to the group template and used to scale the normalized vectors. White matter voxels were determined by segmenting the weighted average template of T1w, T2w, and GFA volumes in FreeSurfer~\citep{FischlSalat2002}. 

\subsection{Estimating voxel expansion due to normalization}

The 3D warps generated by ANTs were used to calculate the Jacobian matrix at each voxel. The log of the determinant of this matrix is an indication of whether the tissue in that voxel expanded or contracted in size in order to match the template images \citep{kim2008structural}. These values are commonly used to test for morphological differences between groups. We use these log-Jacobian values to dismiss any systematic morphological or misregistration effects that might affect this study (see~\ref{sec:appendix_jacobians}).

\subsection{Subjects}

The Human Connectome Project includes 109 pairs of twins, of which 57 are monozygotic (MZ) and 52 are dizygotic (DZ), and a further 47 pairs of non-twin siblings which are disjoint from the population of twins. The analysis in the rest of this paper is focused on the subjects in these three populations. Table~\ref{tab:demographics} details the demographic information of these subjects.

The control groups for the twin and sibling populations are obtained by sampling with replacement an equal number of pairs of non-related subjects from the same population. We control for gender- and age-related confounders by matching gender and age-ranges such that the control populations have the same demographic distribution as detailed in Table~\ref{tab:demographics}. We refer to these control groups of non-related pairs of individuals as \emph{strangers}.

\begin{table*}
\centering
\caption{Age and gender demographics of each pair of twins and siblings in the study population.}
\label{tab:demographics}
\begin{tabular}{lrrrrrr}
\toprule
\begin{tabular}{@{}l@{}}\bf{Monozygotic} \\ \bf{twins}\end{tabular}  &  \specialcell{22-25\\22-25} &  \specialcell{22-25\\ 26-30} &  \specialcell{22-25\\ 31-35} &  \specialcell{26-30\\ 26-30} &  \specialcell{26-30\\ 31-35} &  \specialcell{31-35\\ 31-35} \\
\midrule
Both female & {} & {} & {} & 24 & {} & 19 \\
Both male & 3 & {} & {} & 7 & {} & 4 \\
\midrule
\begin{tabular}{@{}l@{}}\bf{Dizygotic} \\ \bf{twins}\end{tabular} &  {} &  {} &  {} &  {} &  {} &  {} \\
\midrule
Both female & 1 & {} & {} & 17 & {} & 13 \\
Both male & 5 & {} & {} & 9 & {} & 7 \\
\midrule
\textbf{Siblings} &  {} &  {} &  {} &  {} &  {} &  {} \\
\midrule
Both female & 1 & 3 & {} & 1 & 2 & 2 \\
Mixed gender & 6 & 7 & {} & 3 & 7 & 2 \\
Both male & 1 & 4 & 1 & 2 & 3 & 2 \\
\bottomrule
\end{tabular}
\end{table*}

\subsection{Defining voxel-wise similarity}

We seek to identify regions of white matter that contain significantly more similarly oriented white matter structures within a population of interest when compared to a control population. We first present a similarity metric between subjects defined with respect to a single voxel. We then extend that metric to measure similarity, or coherence, across \textit{dyads of neighboring voxels} between pairs of subjects. Connected components of dyads that are significantly similar within the population of interest form the arbitrarily shaped regions of white matter associated with that population.

We define a similarity metric between a pair of subjects on the basis of their 6D ODFs within a voxel. We reduce the distributions down to the most probable underlying oriented white matter microstructure, which we describe as an ordered series of vectors, by extracting local peaks of the ODFs as MDA vectors. We order MDA vectors from the same voxel by their magnitude normalized by the isotrophic component of the distribution. We extract the four largest peaks in any given MDA distribution, as detailed in Section~\ref{sec:mda}.

We measure how similar a pair of individuals are with respect to their oriented white matter microstructure with\-in a voxel by comparing their extracted MDA vectors. The similarity should take into account similarity in both direction and magnitude. Common methods to compare vector similarity that incorporate both magnitude and direction include the dot product as well as various p-norms of the vector differences. We use the $L^2$ norm, or Euclidean distance, to compare individual vectors. This choice of metric corresponds well to the geometric space the measured microstructure exists in, as well as is robust to noisy MDA vectors which manifest as having small magnitudes.

Let $X_v^i$ be the $i$-th three-dimensional vector in voxel $v$ for subject $X$. This is a directed vector representation of an ODF, which fundamentally is not directed (as seen in Fig.~\ref{fig:cartoon}b). As such, when computing a dissimilarity between ODFs we consider the minimum distance between the vector representation of an ODF \emph{or its reflection around the origin} to another such vector representation \emph{or that vector's reflection around the origin}, where the origin is the center of a given ODF. We compute subject $X$'s dissimilarity to subject $Y$ in voxel $v$ as

\begin{equation}
\label{eq:voxel_dissimilarity}
d(X,Y,v) = \sum_{i=1}^k \min\left(\norm{X_v^i - Y_v^i}, \norm{X_v^i + Y_v^i}\right),
\end{equation}

where we have $k$ MDA vectors and use the $L^2$ norm in $d=3$ dimensions,

\begin{equation}
\norm{x} = \sqrt[]{\sum_{k=1}^d |x_k|^2}.
\end{equation}

\subsection{Defining voxel dyad similarity}

As we report in~\ref{sec:appendix_voxel_results}, Eq.~\ref{eq:voxel_dissimilarity} is a suitable meth\-od to discover white matter voxels that are significantly more similar in a population of interest when compared to a control population, and that these voxels spatially cluster into white matter regions. However, we derive a related method that directly encodes the notion of spatially adjacent voxels and serves as a more natural way to discover white matter regions.

Individually significant voxel dyads can be aggregated to form large arbitrarly shaped white matter regions. To this end, we define a lattice network over the white matter voxels and search for subnetworks, or white matter regions, that exhibit significant similarity. Each white matter voxel serves as a node in this network, and we consider dyads of neighboring voxels (those that share a common face, edge, or corner, i.e. each voxel may have up to 26 spatially adjacent white matter voxels that form a cube surrounding the center voxel). This forms a lattice network of the white matter voxels connecting the nearly one million white matter voxels together as 12.2 million voxel dyads. We again exclude those voxels which we have evidence for being more similarly registered in the population of interest, as reported in~\ref{sec:appendix_jacobians}.

Using this network approach, the random variable of interest no longer corresponds to a single voxel but instead to dyads of neighboring voxels. We modify Eq.~\ref{eq:voxel_dissimilarity} such that subjects $X$ and $Y$'s dissimilarity with respect to the undirected voxel dyad $(u, v)$ is

\begin{equation}
\label{eq:sig_edges}
\begin{split}
d(X,Y,u,v) = \frac{1}{2} \sum_{i=1}^k \Big( &\min\left(\norm{X_u^i {-} Y_v^i}, \norm{X_u^i {+} Y_v^i}\right)\\
                                            + &\min\left(\norm{X_v^i {-} Y_u^i}, \norm{X_v^i {+} Y_u^i}\right) \Big) ,
\end{split}
\end{equation}

where we take the arithmetic mean between the directed pairs $(u, v)$ and $(v, u)$ such that the dissimilarity is symmetric between subjects (and reordering of the dyad) as is Eq.~\ref{eq:voxel_dissimilarity}.

This dissimilarity can be considered as the \textit{incoherence} between neighboring voxels in a pair of subjects. A low dissimilarity implies that the fiber populations in the two voxels across subjects contain similarly oriented white matter structures, though not necessarily that there exists a fiber population that travels between the two voxels. A high dissimilarity could correspond to perpendicular fiber populations, or large deviations in the measured magnitude of the MDA peaks. An example of this dissimilarity, or incoherence, can be seen in Fig.~\ref{fig:cartoon}(c).

\subsection{Population differences and significance}

\begin{figure}
\centering
\begin{subfigure}[t]{.24\textwidth}
  \centering
  \includegraphics[width=0.97\textwidth]{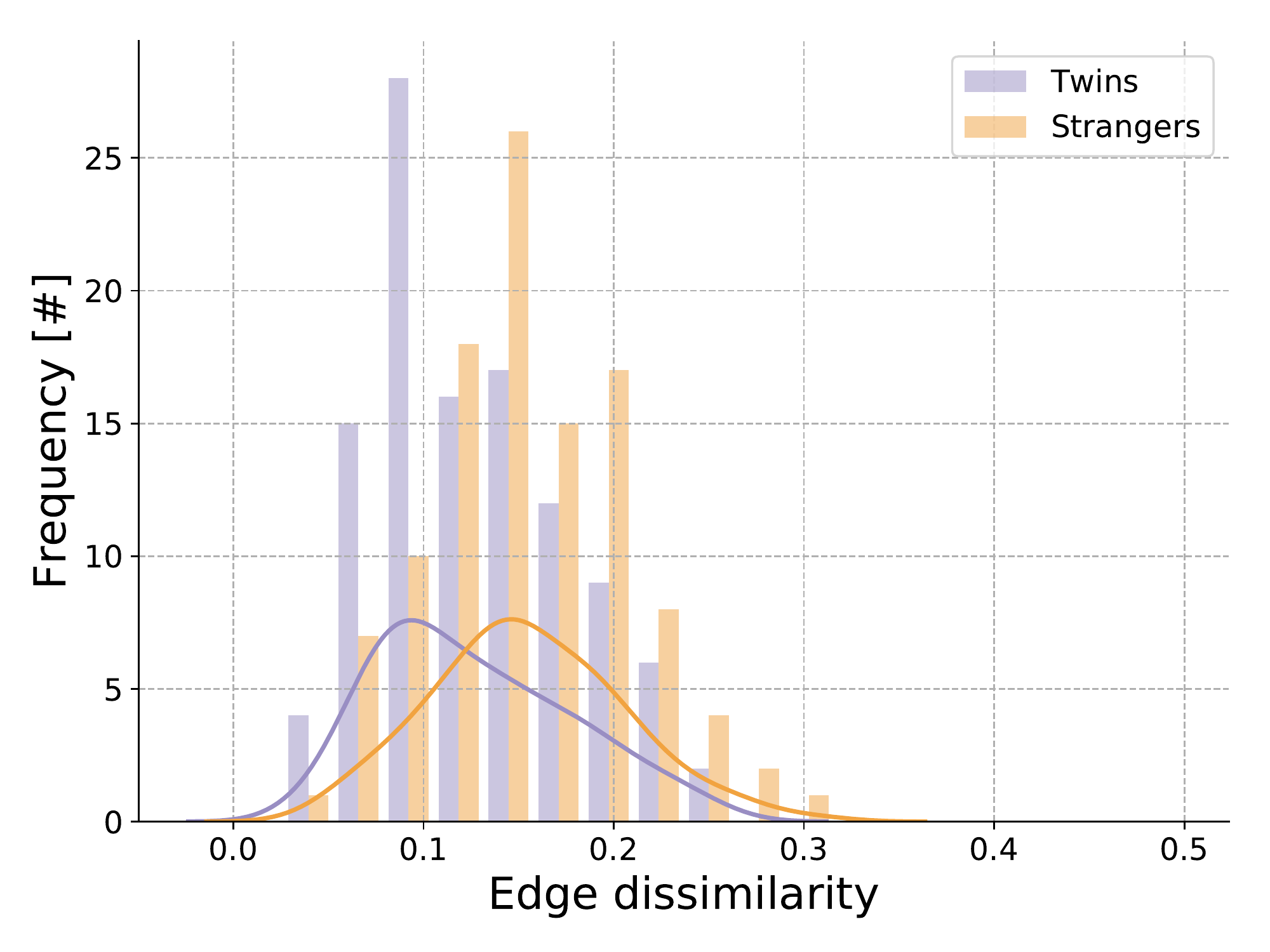}
  \begin{adjustwidth}{5pt}{5pt}
  \caption{Voxel dyad in which twins are significantly more similar than strangers. Kernel density estimates overlaid as solid lines.}
  \label{fig:sub_sig_similar}
  \end{adjustwidth}
\end{subfigure}%
\begin{subfigure}[t]{.24\textwidth}
  \centering
  \includegraphics[width=0.97\textwidth]{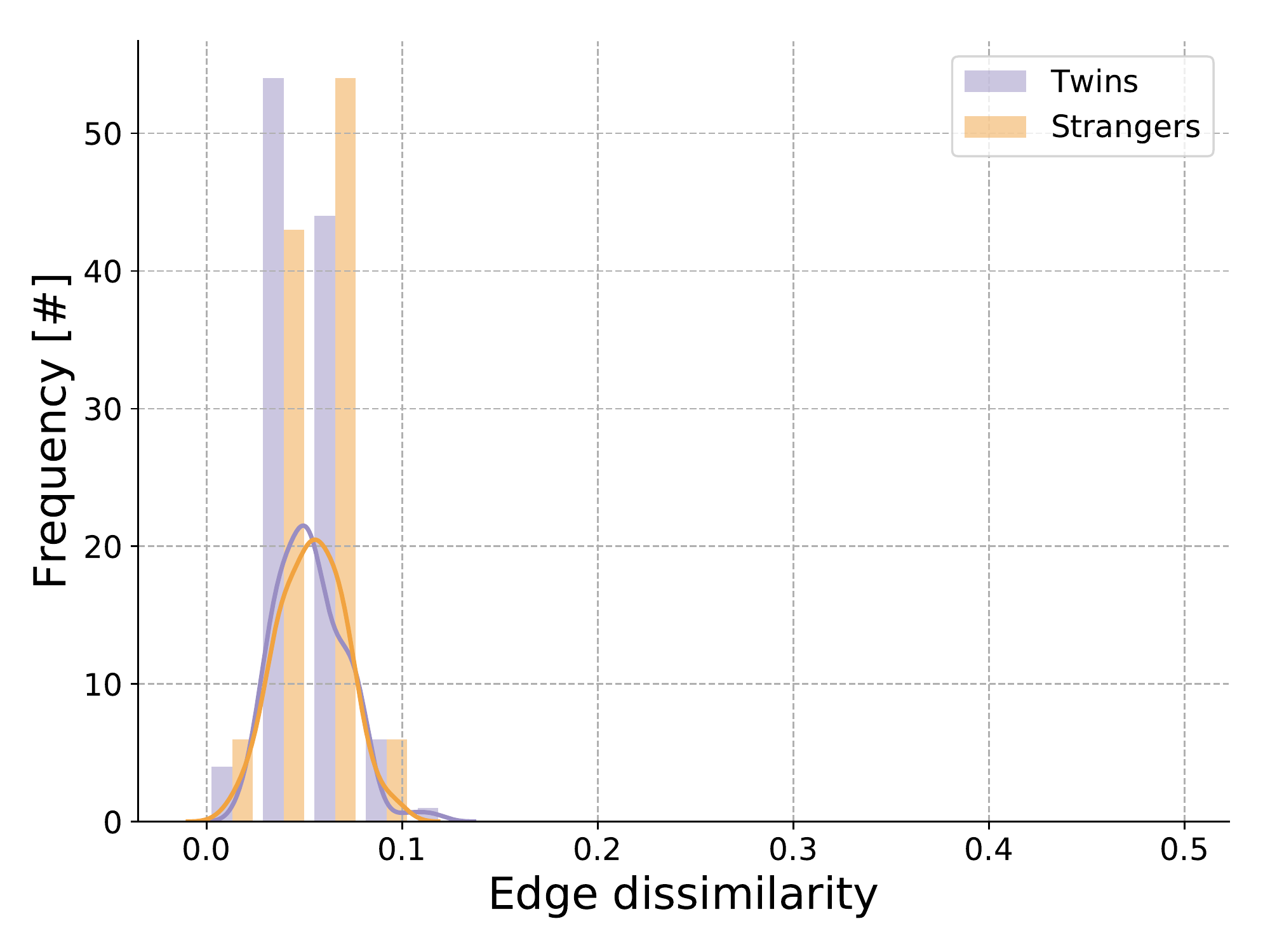}
  \begin{adjustwidth}{5pt}{5pt}
  \caption{Voxel dyad in which twins are \emph{not} significantly more similar than strangers. Kernel density estimates overlaid as solid lines.}
  \label{fig:sub_nonsig_similar}
  \end{adjustwidth}
\end{subfigure}
\caption{Example distributions of similarities as computed from Eq.~\ref{eq:sig_edges} between all twins and all strangers from two dyads, (\ref{fig:sub_sig_similar}) from a dyad that is significantly more similar within twins than strangers and (\ref{fig:sub_nonsig_similar}) from a dyad where no significant differences exist. For clarity, a non-parametric kernel density estimation has been overlaid.}
\label{fig:similarity_example}
\end{figure}

For each dyad $e = (u,v)$ of spatially adjacent voxels $u$ and $v$ in the white matter lattice network we obtain a sample of the distribution of dissimilarities in the population under consideration, $e_1^T, e_2^T, \dots, e_n^T$, and for the control population, $e_1^S, e_2^S, \dots, e_n^S$, empirically:

\begin{equation}
e^T_i = d(T_{i,1}, T_{i,2}, u, v),
\end{equation}
\begin{equation}
e^S_i = d(S_{i,1}, S_{i,2}, u, v).
\end{equation}

Where $T_i$ and $S_i$, $1 \leq i \leq n$, are the $i$-th pair of subjects from the population under consideration and control population, respectively.

We seek to identify those dyads in which the pairs of subjects from the population of interest are significantly more similar than that in the control population. As the distributions of dissimilarities $e^T$ and $e^S$ are non-normal, instead of a t-test we employ a Mann-Whitney U test \citep{mann1947test} to test for differences in the two distributions. Having visually verified that the same shape assumption holds, the Mann-Whitney U test is a non-parametric rank test of the null hypothesis that the two samples of dissimilarities from edge $e$ are equally likely to be as large, 

\begin{equation}
\label{eq:null_hypothesis}
P(v^T < v^S) = P(v^T > v^S),
\end{equation}

against the one-sided alternative hypothesis that the population of interest tends to have lower dissimilarities, 

\begin{equation}
\label{eq:alternative_hypothesis}
P(v^T < v^S) > P(v^T > v^S); 
\end{equation}

that is that the population of interest tends to be more similar. For the dyads in which we reject the null hypothesis, we have evidence that the population of interest is more coherent, or similar. Fig.~\ref{fig:similarity_example} shows an example distribution of coherence from dyads that are and are not significantly similar among a population of interest (twins) as compared to a control group (strangers).

Alternatively, for populations of interest for which the analyst hypothesizes should have common \emph{less coherent} regions\textemdash such as populations of subjects with neurological disorders or injuries\textemdash the analyst might instead test against a one-sided alternative hypothesis that the population of interest tends to have higher dissimilarities. In either case, care must be taken to not aggregate together dyads using a two-sided hypothesis, or dyads from different one-sided hypotheses, as a region formed by such aggregated dyads does not form a single unified region of interest.

To account for multiple hypothesis across all neighboring voxels in white matter, we estimate the false-discovery rate given a particular p-value threshold~\citep{benjamini1995controlling, genovese2002thresholding}. 
Fig.~\ref{fig:edge_pvalues} shows a flat baseline of p-values for this null hypothesis with a sharp peak as $p$ approaches zero indicating high statistical power of the test being employed.

\subsection{Defining regions and region-wise similarity}

Of the set of neighboring white matter voxels for which we reject the null hypothesis and that have been corrected for multiple hypothesis, we further prune unlikely spatially isolated dyads. This is accomplished by aggregating together dyads that form connected components in the white matter lattice network and keeping only the largest such components. These form arbitrarily shaped disjoint \emph{regions} of white matter that can each be considered as single units of interest for further analysis. 

For a pair of subjects $X$ and $Y$ and a white matter region $\mathcal{R}$ (a set of white matter voxel dyads) we defined a dissimilarity between the subjects with respect to $\mathcal{R}$ as the median dissimilarity of all dyads in $\mathcal{R}$ using Eq.~\ref{eq:sig_edges},

\begin{equation}
\label{eq:region_dissimilarity}
d_\mathcal{R}(X, Y) = \text{median} \left( \left\lbrace d(X, Y, u, v), \forall (u,v) \in \mathcal{R} \right\rbrace \right).
\end{equation}

\subsection{Defining between-subject similarity}

We define a single dissimilarity measure between a pair of subjects $X$ and $Y$ on the basis of multiple white matter regions as the mean region similarity across each of the white matter regions $\mathcal{R}_0, \mathcal{R}_1, \ldots, \mathcal{R}_{R-1}$,

\begin{equation}
\label{eq:subject_sim}
d(X,Y) = \frac{1}{R} \sum_{i=0}^{R-1} d_{\mathcal{R}_i}(X,Y).
\end{equation}

\section{Results}
\label{sec:results}

For each dyad of neighboring white matter voxels, we computed the incoherence using Eq.~\ref{eq:sig_edges} with $k=1$ MDA peak across each pair of subjects in the monozygotic (MZ) twin, dizygotic (DZ) twin, and matched stranger populations. We found those white matter dyads for which we have enough evidence to rule out the null hypothesis described by Eq.~\ref{eq:null_hypothesis} in favor of the alternative hypothesis given by Eq.~\ref{eq:alternative_hypothesis}, where we consider the population of interest all pairs of MZ and DZ twins (and \emph{not} the non-twin sibling pairs). We then examined the largest connected subnetworks and their properties.

We control for similarity due to the morphology of the brain that would otherwise confound this analysis by excluding voxels which can be shown to have been similarly morphed into the normal space in twins but not in strangers (see~\ref{sec:appendix_jacobians}).

\subsection{Twin-similar white-matter regions}
 
\begin{figure}
\centering
\includegraphics[width=0.35\textwidth]{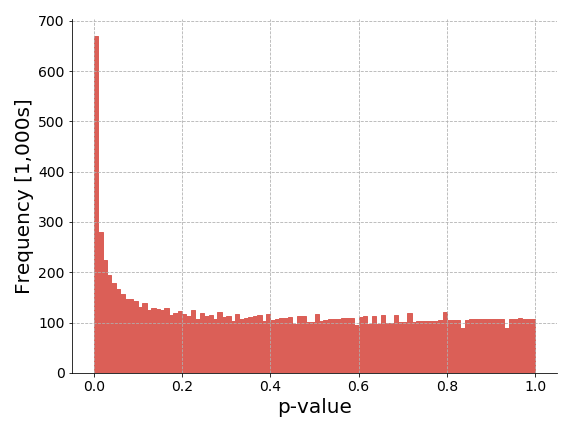}
\caption{Distribution of p-values for each dyad of neighboring white matter voxels assuming the null hypothesis in Eq.~\ref{eq:null_hypothesis}, that monozygotic and dizygotic twins are not more similar than strangers.}
\label{fig:edge_pvalues}
\end{figure}

\begin{figure}
\centering
\includegraphics[width=0.45\textwidth]{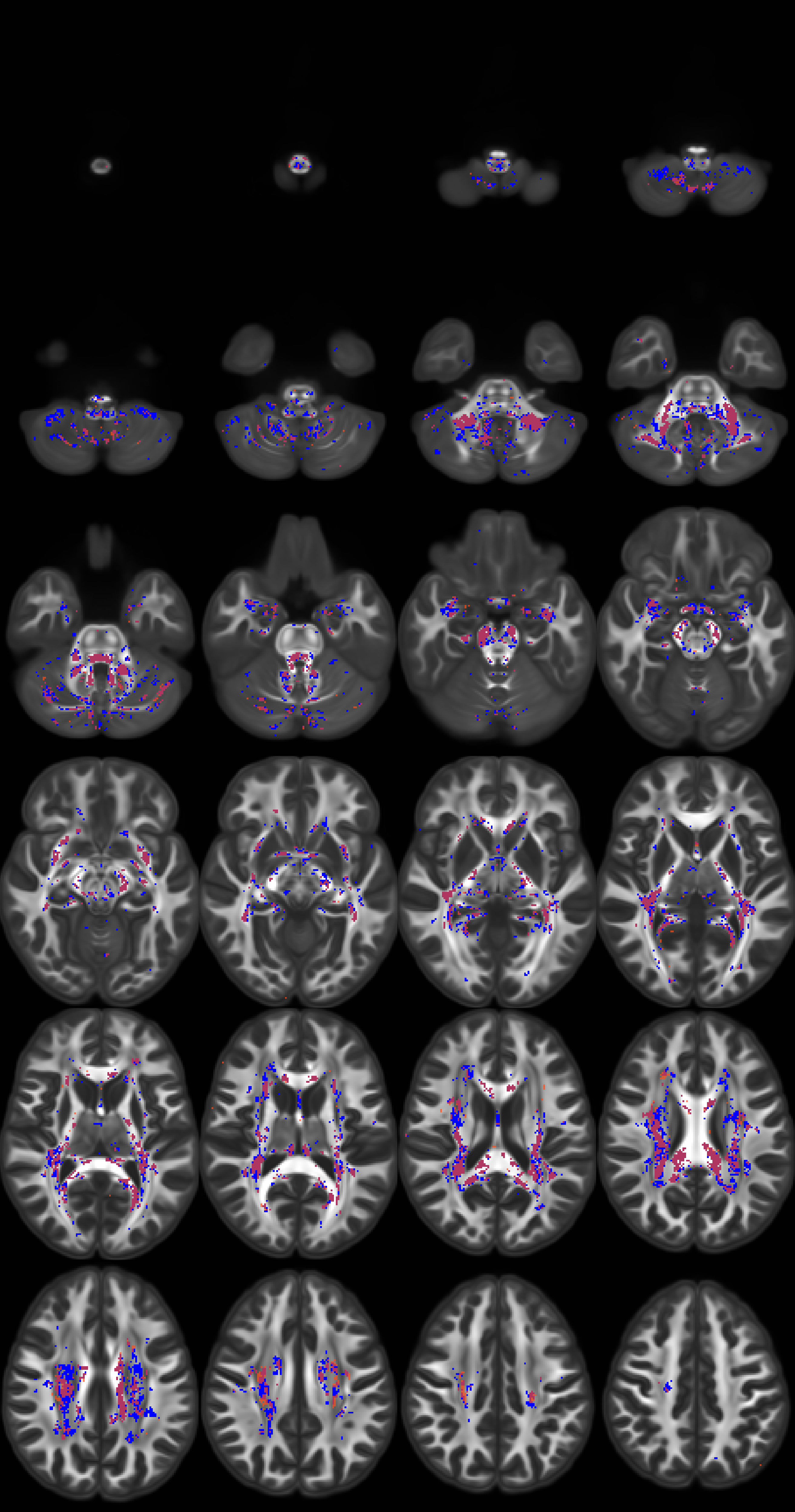}
q\caption{Axial slices of all 35.1k voxels (blue) and 19.3k (red) that were part of a neighboring voxel dyad found to be significantly more similar ($p<10^{-4}$, $\text{FDR} = 1.5\%$) ($p<10^{-4}$, $\text{FDR} = 3.2\%$) among monozygotic and dizygotic twins as compared to a control population of strangers using Eq.~\ref{eq:sig_edges} with 1 and 2 peaks, respectively. Purple voxels are those that feature in the intersection of both, and form the vast majority of the extent of otherwise red voxels. Generalized Fractional Anisotrophy (GFA) template as background. Image created using ITK-SNAP~\citep{py06nimg}.}
\label{fig:all_sig_edges}
\end{figure}

\begin{figure*}
\centering
  \begin{subfigure}[t]{0.48\linewidth}
	\centering
    \includegraphics[height=2.7in]{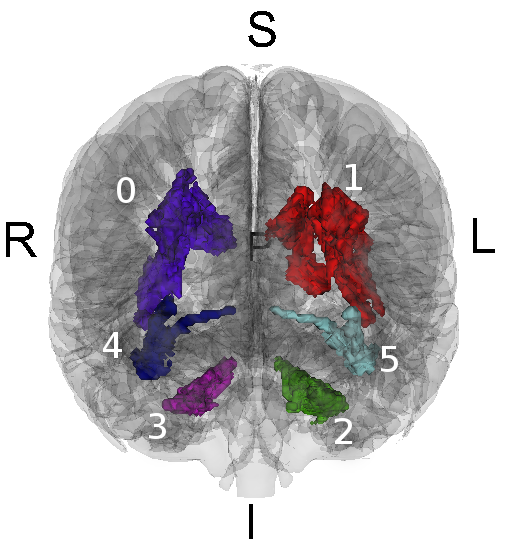}
	\caption{Anterior view of the six largest regions.}
  \end{subfigure}%
  ~
  \begin{subfigure}[t]{0.48\linewidth}
	\centering
    \includegraphics[height=2.7in]{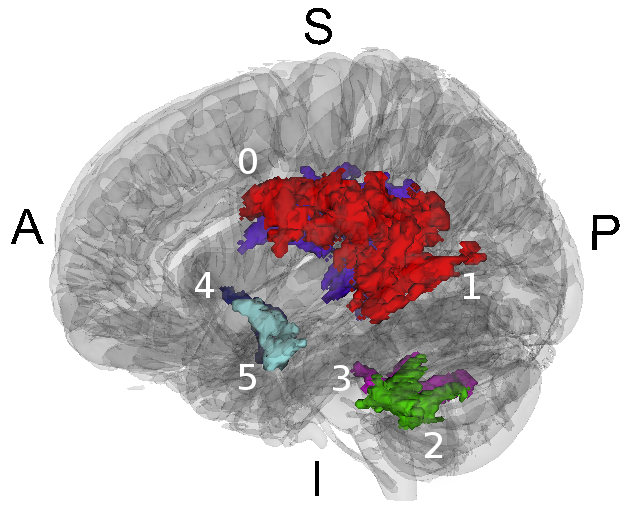}
	\caption{Left view of the six largest regions.}
  \end{subfigure}
 
  \begin{subfigure}[t]{0.48\linewidth}
	\centering
    \includegraphics[height=2.7in]{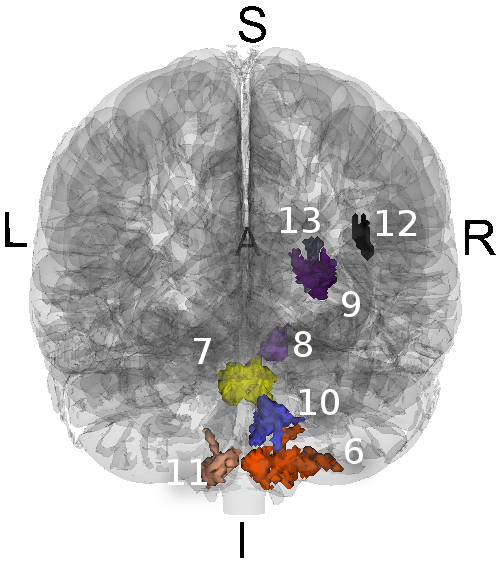}
	\caption{Posterior view of the seventh to 14th largest regions.}
  \end{subfigure}%
  ~
  \begin{subfigure}[t]{0.48\linewidth}
	\centering
    \includegraphics[height=2.7in]{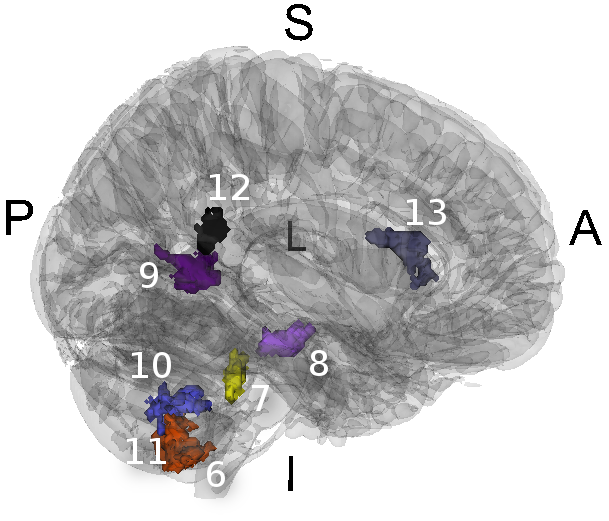}
	\caption{Right view of the seventh to 14th largest regions.}
  \end{subfigure}
 
    \begin{subfigure}[t]{0.48\linewidth}
	\centering
    \includegraphics[height=2.7in]{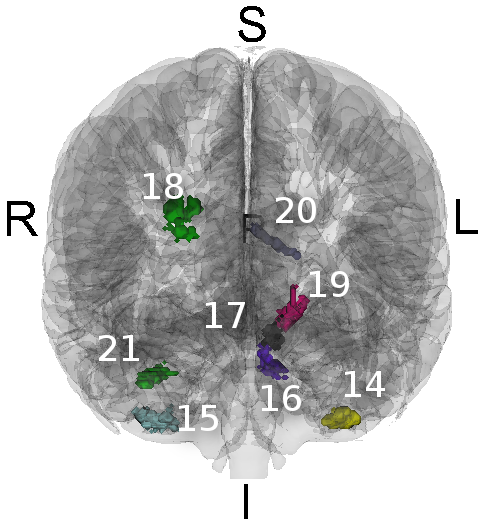}
	\caption{Anterior view of the 15th to 22nd largest regions.}
  \end{subfigure}%
  ~
  \begin{subfigure}[t]{0.48\linewidth}
	\centering
    \includegraphics[height=2.7in]{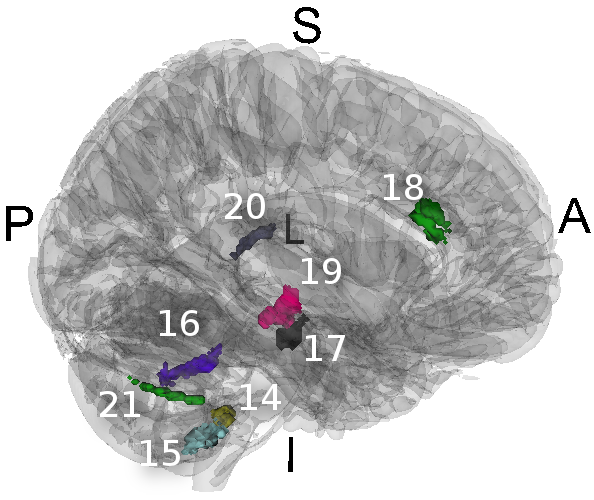}
	\caption{Right view of the 15th to 22nd largest regions.}
  \end{subfigure}
  \caption{The twenty-two largest white matter regions in which monozygotic and dizygotic twins are more similar than a control population of strangers, as visualized on a transparent background of a T1w volume. Image captured using Slicer 4~\citep{fedorov20123d}.}
  \label{fig:top20}
\end{figure*}

Of the 12.2 million dyads of spatially adjacent white matter voxels, we identified 71,857 as significantly more similar within MZ and DZ twins as compared to a matched control group of strangers ($p < 10^{-4}$, false discovery rate 1.5\%), see Fig.~\ref{fig:edge_pvalues}. These dyads contained 35,119 unique white matter voxels, as seen in Fig.~\ref{fig:all_sig_edges}. The dyads form 3,145 connected components in the white matter lattice network, of which 1,791---more than half---were trivial subnetworks of a single dyad of two voxels. More interestingly, twenty-nine subnetworks connected more than one hundred voxels, or a volume of white matter that is approximately $200 \text{mm}^3$ in normalized template space. We selected the twenty-two largest subnetworks as units for further analysis as these comprise 75\% of all significant voxel dyads. See table~\ref{tab:region_info} for relevant statistics of these twenty-two largest white matter regions, and Fig.~\ref{fig:top20} for a visualization of these twenty-two white matter regions.

Increasing the number of peaks from $k=1$ in Eq.~\ref{eq:sig_edges} steadily decreases the size and significance of the results. With $k=2$ peaks, we observe less than half the significant voxel dyads or 35,268 dyads containing 19,253 unique voxels ($p < 10^{-4}$, FDR 3.2\%) and which can be seen to be nearly fully encompassed by the results with $k=1$ in Fig.~\ref{fig:all_sig_edges}. The largest cluster of voxels identified using two peaks and not one is in the centrum semiovale, which has previously been identified as an area containing multiple crossing fibers~\citep{Volz2017}. Further increasing to $k=3$ peaks reduces the significant dyads to 30,134 containing 15,623 unique voxels ($p < 10^{-4}$, FDR 4.0\%). This decrease in the size and significance of the results can be attributed to the noise associated with the higher MDA peaks; 32.9\% of voxels are known to be singly connected such that in these the local MDA peaks past the first contain no true signal~\citep{Volz2017}. Unless otherwise stated, all further results are presented with $k=1$ MDA peaks.

Furthermore, significant voxel dyads are biased towards non-diagonal connections between voxels. Forming a lattice network only between voxels which share either a face or edge, allowing up to eighteen neighbors per voxel, and retesting the null hypothesis we discover 58,159 voxel dyads containing 32,986 unique voxels to be significant ($p < 10^{-4}$, FDR 1.2\%). As this reduced lattice network has only 69\% (roughly 18/26) of the dyads present as compared to the original, if no bias towards non-diagonal connections existed we would expect 69\% of the original dyads, or 49.6k, to remain significant. This is much less than what we observe. To some degree, this is not surprising as the diagonal dyads should be expected to be less spatially coherent due to the greater distance between their centers and the smaller effect of spatial smoothing inherent to the imaging process. The aggregate white matter regions remain fairly invariant to the choice of lattice network, with the total number of regions decreasing from 5,699 of which 29 contain at least 100 voxels to 5,310 of which at least 27 contain at least 100 voxels when going from a 26- to 18-connected lattice.
 
\subsection{Effect size of white matter regions}
 
Having aggregated together individually significant vox\-el dyads to form large arbitrarily shaped white matter regions, we measure similarity between pairs of subjects on the basis of a single region using Eq.~\ref{eq:region_dissimilarity}. We show that this region-wise similarity measure corresponds to a large effect size when comparing pairs of MZ and DZ twins to a control group of strangers, and that this measure generalizes to a previously unseen group of siblings. The sibling data was not considered previous to this point with one exception: four siblings were among the 38 subjects sampled from all HCP scans to define a normal template.
 
\begin{figure}
\centering
\includegraphics[width=0.45\textwidth]{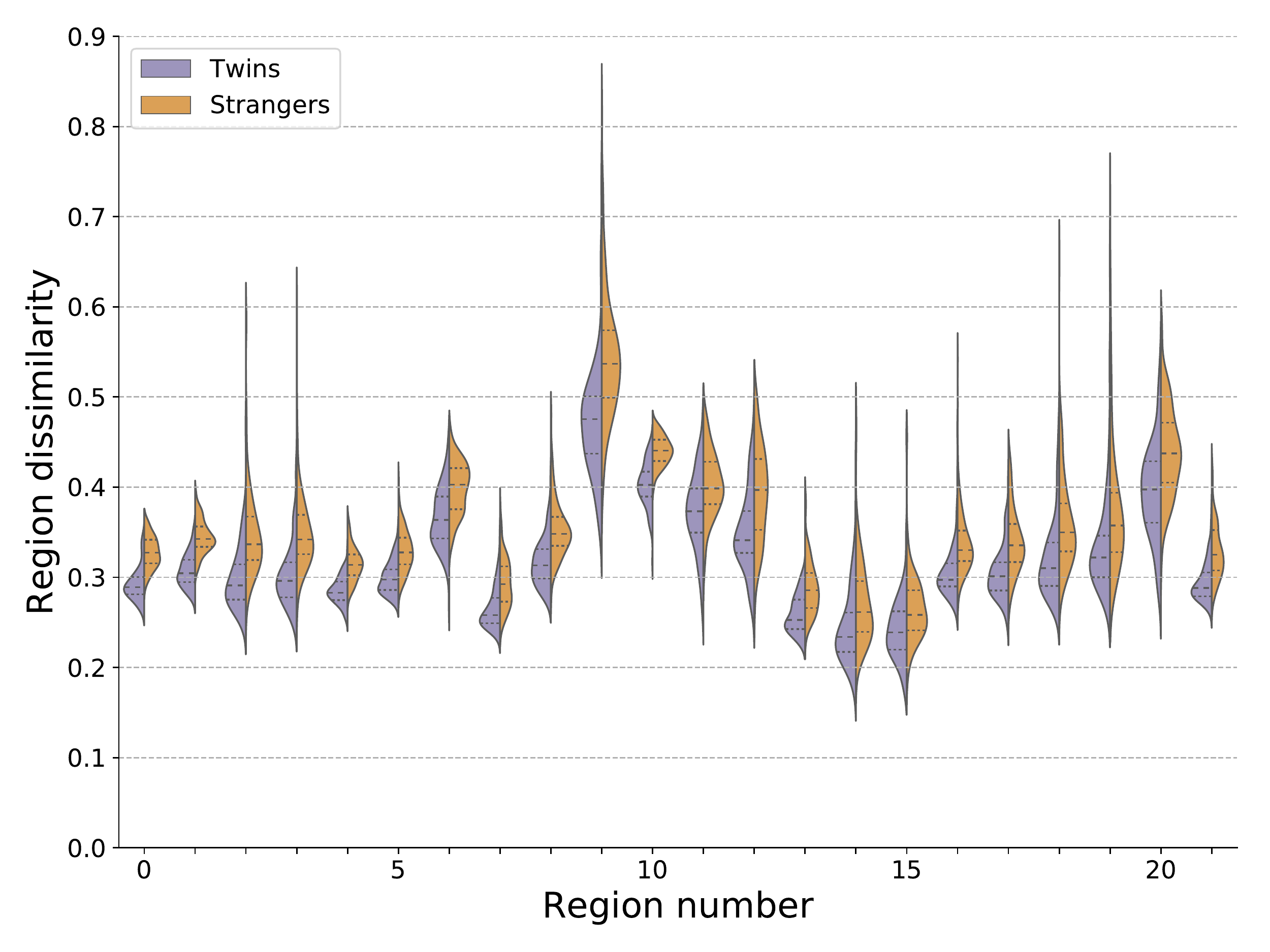}
\caption{Distributions of region dissimilarities $d_{\mathcal{R}}(X,Y)$ per discovered white matter region of the monozygotic and dizygotic twin (left halves) and stranger populations (right halves). Quartiles of each distribution are shown as dashed lines.}
\label{fig:region_dissim}
\end{figure}

In each of the twenty-two largest discovered regions, the distribution of such region dissimilarity in the twin population and the stranger population (as seen in Fig.~\ref{fig:region_dissim}) is, although overlapping, reasonably well separable. Table~\ref{tab:region_info} shows the aggregate statistics of the regions. Larger regions correlate with larger effect sizes (as measured by the Cohen's $d$ statistic), which is to be expected as they are composed of a greater amount of significant voxel dyads, though all regions are associated with very large effect sizes in the range of 0.6 to 2.1 pooled standard deviations. We note that the effect sizes generalize to a population of siblings, which were not used in identifying the regions, showing that this measure of similarity between MZ and DZ twins generalizes to similarity among siblings though with medium effect sizes.
 
\begin{table*}
\caption{White matter regions discovered that are significantly more similar in monozygotic and dizygotic twins than in strangers. Effect size is Cohen's $d$ as compared to the control population of strangers, or difference in means standardized by pooled standard deviations.}
\label{tab:region_info}
\centering
\begin{tabular}{lrrrr}
\toprule
\begin{tabular}{@{}c@{}}\bf{Region} \\ \bf{number}\end{tabular} &  \begin{tabular}{@{}c@{}}\bf{Number} \\ \bf{of dyads}\end{tabular} &  \begin{tabular}{@{}c@{}}\bf{Number} \\ \bf{of voxels}\end{tabular} &  \begin{tabular}{@{}c@{}}\bf{Twin effect} \\ \bf{size [STD]} \end{tabular} &  \begin{tabular}{@{}c@{}}\bf{Sibling effect} \\ \bf{size [STD]} \end{tabular} \\
\midrule
~0  &      17,336 &       5,699 &         1.94 &            0.84 \\
~1  &      14,382 &       5,369 &         2.06 &            0.92 \\
~2  &       5,458 &       1,311 &         1.15 &            0.33 \\
~3  &       3,793 &       1,146 &         1.20 &            0.34 \\
~4  &       1,689 &        745 &         1.74 &            0.82 \\
~5  &       1,374 &        599 &         1.46 &            0.77 \\
~6  &       1,322 &        594 &         1.09 &            0.35 \\
~7  &       1,218 &        431 &         1.13 &            0.66 \\
~8  &        896  &        306 &         1.24 &            0.55 \\
~9  &        874  &        308 &         1.11 &            0.55 \\
10  &        750  &        408 &         1.78 &            0.78 \\
11  &        579  &        222 &         0.77 &            0.25 \\
12  &        576  &        142 &         0.86 &            0.32 \\
13  &        516  &        255 &         1.16 &            0.57 \\
14  &        476  &        156 &         0.58 &            0.39 \\
15  &        420  &        220 &         0.60 &           -0.13 \\
16  &        368  &        173 &         1.02 &            0.29 \\
17  &        340  &        121 &         1.11 &            0.56 \\
18  &        334  &        198 &         0.85 &            0.02 \\
19  &        310  &        184 &         0.63 &            0.29 \\
20  &        308  &        110 &         0.88 &            0.45 \\
21  &        302  &        146 &         1.27 &            0.43 \\
\bottomrule
\end{tabular}
\end{table*}
 
\begin{figure}
\centering
\includegraphics[width=0.45\textwidth]{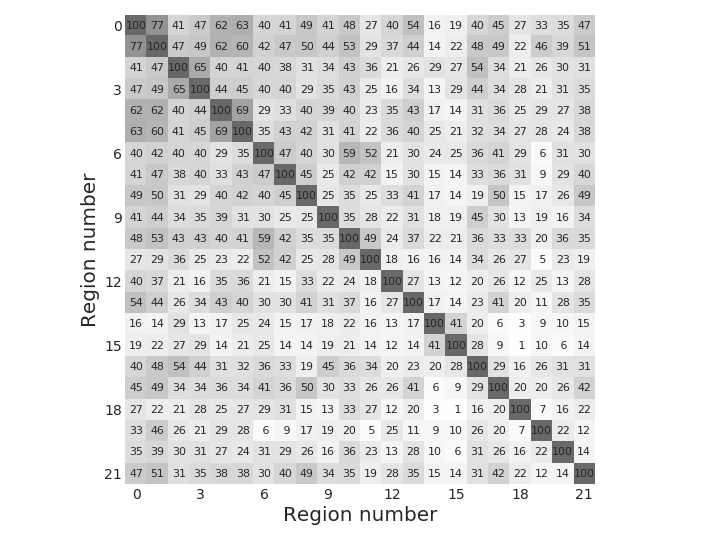}
\caption{ Pearson correlation coefficients $r$ of region dissimilarities $d_{\mathcal{R}}(X,Y)$ of each pair of monozygotic twins, dizygotic twins, siblings, and strangers, for each pair of regions. The percentage correlation is reported as a whole number (i.e., $100r$), with proportional shading added for clarity.}
\label{fig:region_hamming}
\end{figure}

Though these white matter regions all exhibit a large difference in the distributions of MZ and DZ twins and strangers, and to a lesser extent also between siblings and strangers, they do so somewhat independently. We compute the Pearson correlation coefficient between each pair of regions with respect to the measured region similarity for each pair of monozygotic and dizygotic twins as well as strangers, as seen in Fig.~\ref{fig:region_hamming}. We see that larger regions tend to correlate more, which is unsurprising as they discriminate between the groups better. Similarly, regions 14 and 15\textemdash which have the lowest effect sizes\textemdash correlate with each other to a greater extent than with all other regions. There is weak evidence for other such clusters of white matter regions that predict similarly for each pair of subjects. In particular, region pairs that appear mirrored across hemispheres correlate with each other more than they do other regions, such as 4 and 5 or 14 and 15.

\subsection{Subject similarity}
 
\begin{figure}
\centering
	\includegraphics[width=0.45\textwidth]{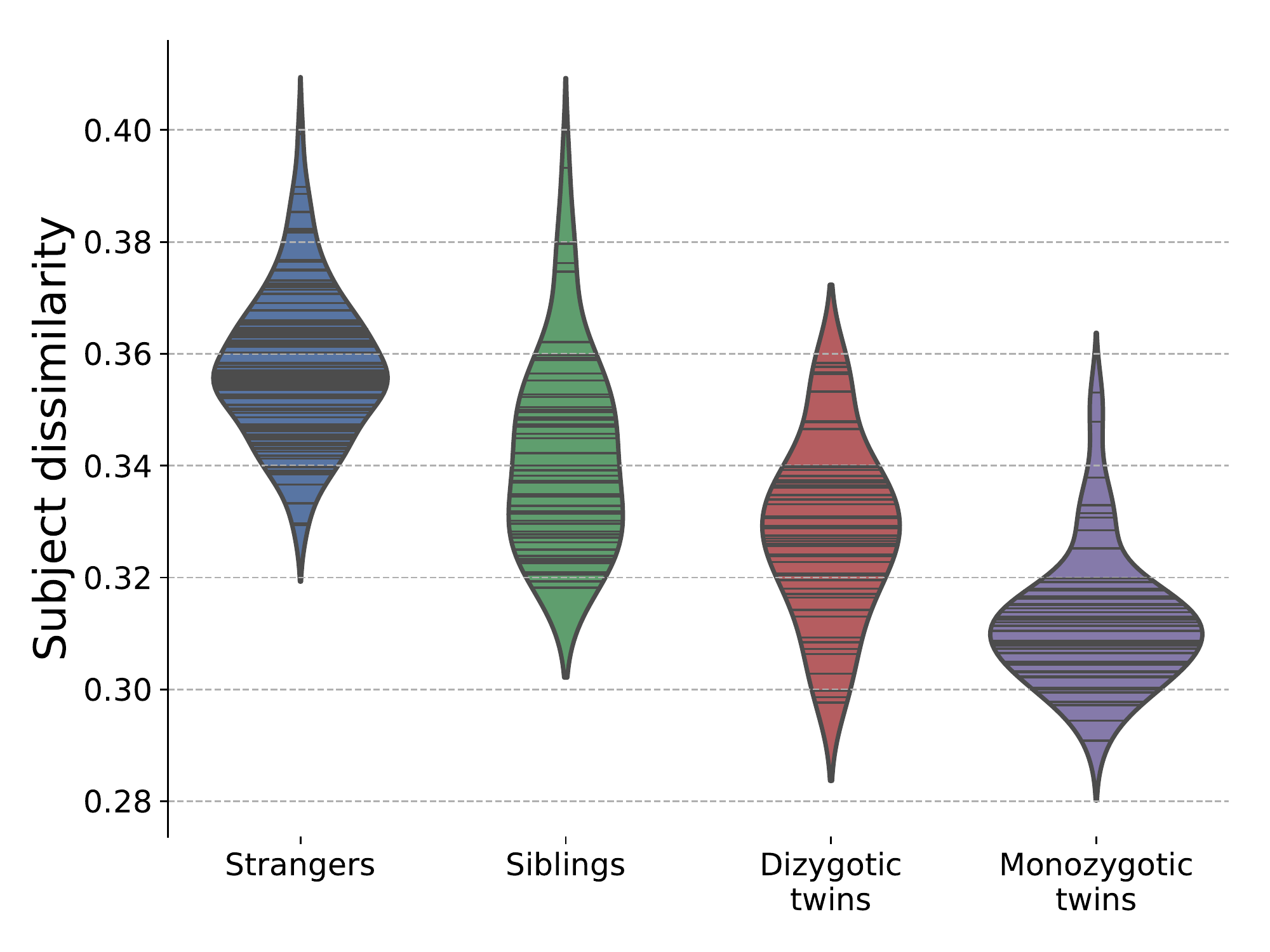}
\caption{The distribution of subject pair dissimilarities $d(X,Y)$ as computed by Eq.~\ref{eq:subject_sim}. Black lines indicate each individual pair of subjects.}
\label{fig:group_predictions}
\end{figure}

Using Eq.~\ref{eq:subject_sim} we measure a single dissimilarity between each pair of MZ twins, DZ twins, non-twin siblings, and strangers on the basis of the $R=22$ discovered white matter regions. The distribution of dissimilarities for each of these groups can be seen in Fig.~\ref{fig:group_predictions}. The modes of the distributions define a clear order from least to most genetic similarity with strangers having high dissimilarities and MZ twins very low, with DZ twins situated in-between. We see this measure generalizes to the population of non-twin siblings, whose data were not used in obtaining the regions, and which have comparable dissimilarities to DZ twins though with a longer tail of high dissimilarities. The majority of this long tail corresponds to mixed-gender sibling pairs, of which none exist in the DZ twin population. Aside from several of the mixed-gender sibling pairs in the long-tail, the majority of mixed-gender sibling pairs are also well separable from the stranger population.
  
\section{Discussion}
\label{sec:discussion}

We have identified a large fraction of deep white matter as being associated with genetic similarity. Areas of white matter with genetic similarity include the superior longitudinal fasciculus, the optic radiations, the middle cerebellar peduncle (particularly near the cerebellar nuclei), the corticospinal tract (through the posterior limb of the internal capsule and cerebral peduncle), and within the anterior temporal lobe adjacent to the amygdalae. 
These regions encapsulate nearly all of deep white matter. 
The large spatial extent of similarity among twins may reflect how fascicles are spatially arranged during neonatal development. Indeed, similarity of deep white matter organization may be partially responsible for similarity in gray matter thickness and curvature if Van Essen's tension hypothesis is true~\citep{van1997tension}. 

Previous voxel-based studies of white matter associated with genetic similarity have identified overlap with our results. For example, \cite{jahanshad2010genetic} report that voxels in the temporal and frontal lobe subregions as showing the highest genetic influence, and additionally they identify significant subregions underlying posterior cortex. \cite{chiang2011genetics} report large portions of white matter as being affected by genetic control and the interaction of that with age, sex, socioeconomic status, and intelligence quotient. Both of these studies focused on per-voxel fractional anisotrophy scalars. In contrast, \cite{chiang2011genetics} consider orientation in a limited manner and report heritable effects in deep white matter based on the amplitudes of the first and second peaks of the per-voxel fiber orientation distribution.

We see a greater effect of white matter similarity with increasing genetic similarity. Monozygotic (MZ) twins are consistently more similar as compared to dizygotic (DZ) twins, which also display greater variance in their similarity. Non-twin siblings likewise are shown to be as similar in terms of these twenty-two white matter regions as DZ twins, though a longer tail of less similar pairs exists that is on par with strangers. This long tail is nearly entirely composed of mixed-gender siblings, however. No mixed-gender sibling exists in either of the MZ or DZ twin groups.

That these results generalize to non-twin siblings serves as an important validation of our method and results. These non-twin siblings were not used in the derivation of the twenty-two white matter regions except for being including as a larger group of Human Connectome Subjects to obtain a normalized template space from which the diffusion voxels were compared in.

The voxel dyads which we discover to be associated with genetic similarity are ones in which there exists sufficient individual variability for there to exist a group-wise difference, and which display a strong enough similarity across most twins as compared to the non-related strangers. There is a strong assumption made that all pairs of interest are similar in the same way, i.e. that a single model of genetic similarity in white matter is sufficient to describe differences between all related siblings and non-related strangers. A promising future direction for this work is to consider multiple effects of similarity to exist in the population under consideration, such as looking for (potentially non-disjoint) partitions of the population such that the each has a strong similarity within only a single or small number of regions.

A potential limitation of the spatial coherence approach is in areas of high curvature. Such areas exhibit rapid changes in orientation between adjacent voxels and as such naturally have a lower baseline of spatial coherence. Though it is still possible to expect a population of interest to have a greater coherence in such areas, it might be harder to pick out the effect statistically which in turns biases results away from such regions.

In our results, we see that the effect of increasing the number of MDA peaks $k$ above one decreases the extent and significance of the voxels identified. It is reasonable to assume this is because the additional orientations are largely associated with noise as they are in white matter regions where it is unlikely there are crossing fibers (per \cite{Volz2017}, a study on the same dataset, about 32.9\% of white matter voxels are singly connected). Indeed, in Fig.~\ref{fig:all_sig_edges}, we see all major areas identified as significantly similar among twins with $k=2$ peaks as being encompassed by the result with $k=1$ peak with an exception of the centrum semiovale which has previously been identified as an area containing multiple crossing fibers. As we do not observe new large areas outside of the intersection of these two results, we conclude that either the similarity in the first peak is sufficient to detect regions associated with this population or that we are not adequately able to incorporate multiple orientations in our method in a manner that is robust enough against noise.

These twenty-two white matter regions, due to being associated with genetic background and/or upbringing common to the twins and siblings, could serve as a first approximation for a basis of defining white matter fingerprints that could be used to identify an individual over the course of their lifetime~\citep{yeh2016quantifying}. However, this analysis might overlook regions that would be better suited to fingerprint individuals that are not preserved between pairs of twins or siblings, i.e. are not associated with genetic similarity but instead some broader concept of individual variance. A promising future direction for this method is in the application to a population of pairs of scans obtained from the same individual for a set of subjects, especially over a period of years, so as to understand what white matter regions contribute to such individual variance and how that changes over the course of our lifetimes.

A crucial component to the presented method is that it considers \emph{pairs} of subjects as the fundamental unit of analysis, and not a single subject. This has two immediate consequences. The first is that it narrows down the scope of the analysis to those regions of white matter which display high similarity, or a small distance for some measure of distance, which is amenable to statistical analysis. The second is that this expands the size of the population under consideration from $N$ individual scans to the order of $N^2$, given that sufficient care is taken during the analysis in sampling pairs and interpreting results, and given that every subject can be expected to have a high pair-wise similarity to the rest of the population of interest (which is not the case for the twin and sibling populations considered in the results). This aspect of considering pairs of input data is akin to Siamese Neural Networks~\citep{koch2015siamese}, which have achieved state-of-the-art performance for learning models with very limited data and which has previously been applied to clinical diagnosis from functional MRI data~\citep{ktena2017distance}.

\section{Conclusion}

In this work we presented a method for identifying spatially contiguous but otherwise arbitrarily shaped white matter regions that are associated with a population of interest. This is a bottom-up approach which builds on the simplest possible building block, or neighboring white matter voxel dyads. We defined a similarity metric on such dyads and find a subset which are significantly more similar within the population of interest as compared to a control population, and control for multiple hypothesis testing. The largest such regions, composed of maximally sized overlapping dyads, are used for further analysis: a region-wise similarity is defined and is shown to have a large effect size between the two populations and generalizes to a previously unseen portion of the population of interest. Finally, a single similarity between a pair of subjects is defined on the basis of a set of such regions and is shown to separate the populations well.

This method is demonstrated on a population of mono\-zy\-got\-ic (MZ) and dizygotic (DZ) twins, with a control group composed of the same individuals with their pairings scrambled such as to keep the same demographic profiles but otherwise form unrelated strangers. The method discovers  3.7\% of all white matter voxels to be associated with genetic similarity (35.1k voxels, $p < 10^{-4}$, false discovery rate 1.5\%), 75\% of which form twenty-two contiguous white matter regions. These white matter regions generalize to a population of non-twin siblings and are shown to be a good indicator of genetic similarity there as well, as compared to a population of strangers. The regions encapsulate nearly all of deep white matter.

\section{Acknowledgements}
This research was supported by a Head Health Challenge grant from General-Electric and the National Football League and Institute for Collaborative Biotechnologies through grant W911NF-09-0001 from the U.S. Army Research Office.

\bibliography{bibliography.bib}

\appendix
\section{Voxel-wise population differences}
\label{sec:appendix_voxel_results}

The results presented in Section~\ref{sec:results} follow from the application of Eq.~\ref{eq:sig_edges}, which is an extension of Eq.~\ref{eq:voxel_dissimilarity} to two neighboring voxels. However, Eq.~\ref{eq:voxel_dissimilarity} is of independent interest and can be used to show similar results as Eq.~\ref{eq:sig_edges}.  

\begin{figure}[ht]
\centering
\includegraphics[width=0.49\textwidth]{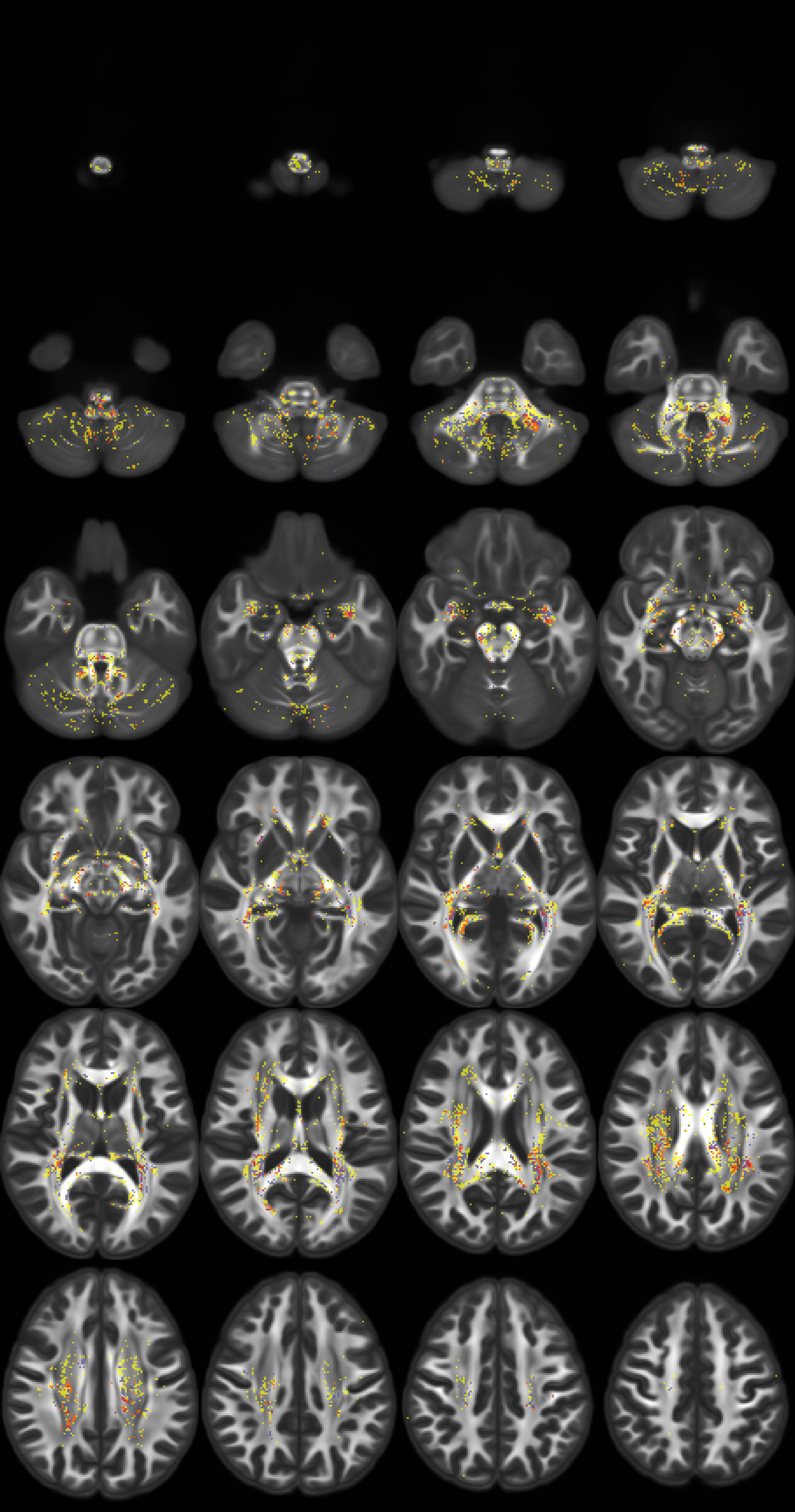}
\caption{Axial slices of voxels identified as significantly more similar within twins than strangers (yellow), and the subset of those voxels found to be significantly more similar within monozygotic compared to dizygotic twins (teal) and in siblings compared to strangers (orange), as computed using Eq.~\ref{eq:voxel_dissimilarity}. Background image is of a population averaged Generalized Fractional Anisotrophy (GFA), where lighter regions indicate higher GFA values. Image created partly using ITK-SNAP~\citep{py06nimg}.}
\label{fig:heritable_results}
\end{figure}

Of the nearly one million white matter voxels we identify 33,003 voxels as significantly more similar within mono\-zy\-got\-ic (MZ) and dizygotic (DZ) twins than a matched control group of strangers ($p < 10^{-3}$, false discovery rate 1.3\%) using Eq.~\ref{eq:voxel_dissimilarity} with $k=1$ MDA peaks. These voxels are visualized in figure~\ref{fig:heritable_results}. Most of these voxels, 23,566, overlap with the previous results in Section~\ref{sec:results}.

The voxels identified were further used to study similarities within MZ twins as compared to DZ twins and also in siblings as compared to a matched control group of strangers. Of the 33.0k voxels discovered to be more similar within twins than strangers we found 8,746 voxels that were significantly more similar within MZ than in DZ twins ($p$~\textless~.05, FDR 8.7\%), and 5,244 voxels that were significantly more similar in siblings than in strangers ($p$~\textless~.05, FDR 19.2\%). These two results overlapped to a small extent, or a total of 1,637 voxels. 

We limit the null hypothesis space to only those 33.0k voxels discovered to be significant previously as a whole-brain study did not identify any reasonable result with sufficiently low false discovery rate. We note that due to the reduced hypothesis space these results still assume a single model of similarity associated with genetically related pairs. We further note that these results are associated with much less statistical significance, indicating that this model does not generalize as well as that of Eq.~\ref{eq:sig_edges} as seen in Fig.~\ref{fig:group_predictions}. In addition, the small intersection of the results is evidence for a need to model heterogeneity within the population, as mentioned in Section~\ref{sec:discussion}.

The microstructural orientation of the MDA distributions is important to distinguish between twins and the control population of strangers; the magnitude by itself only accounts for a small portion of these results. Repeating the experiment with a modification to Eq.~\ref{eq:voxel_dissimilarity} such that it only considers magnitude and not direction discovers 11,948 voxels as significantly more similar within MZ and DZ twins than in the control group ($p < 10^{-3}$, FDR 4.3\%), of which only 6,235 were also identified as significantly similar in the previous test which included orientation data. These magnitude-only results are clustered in the corpus callosum and near the amygdalae.

Comparing the voxels identified as significant when applying Eq.~\ref{eq:voxel_dissimilarity} to Eq.~\ref{eq:sig_edges} is complicated by the different number of hypothesis when considering voxels and voxel dyads in the two equations, respectively. An alternative middle ground between the two that considers within-voxel differences between subjects, as Eq.~\ref{eq:voxel_dissimilarity} does, but for voxel dyads, as Eq.~\ref{eq:sig_edges} considers:

\begin{equation}
\label{eq:within_voxels}
\begin{split}
d(X,Y,u,v) {=} \frac{1}{2} \sum_{i=1}^k \Big(\!&\min\left(\norm{X_u^i {-} Y_u^i}, \norm{X_u^i {+} Y_u^i}\right)\\
                                            + &\min\left(\norm{X_v^i {-} Y_v^i}, \norm{X_v^i {+} Y_v^i}\right)\!\Big)
\end{split}
\end{equation}

\begin{figure}[ht]
\centering
\includegraphics[width=0.49\textwidth]{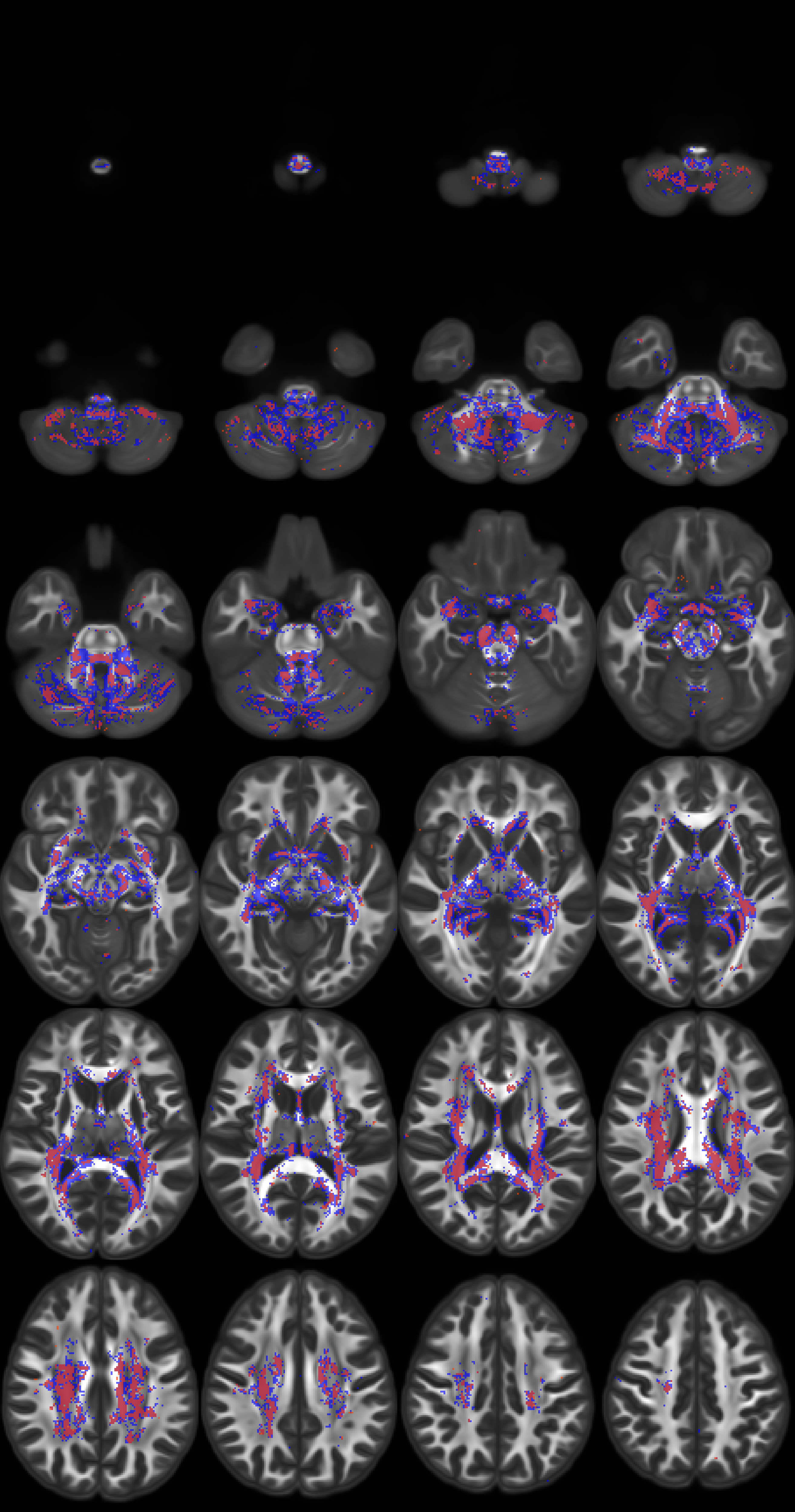}
\caption{Axial slices of voxels identified as significantly more similar within monozygotic and dizygotic twins than strangers as computed using Eq.~\ref{eq:within_voxels} (blue) and Eq.~\ref{eq:sig_edges} (red). Background image is of a population averaged Generalized Fractional Anisotrophy (GFA), where lighter regions indicate higher GFA values. Image created partly using ITK-SNAP~\citep{py06nimg}.}
\label{fig:within_voxels}
\end{figure}

This differs from Eq.~\ref{eq:sig_edges} by only comparing subjects $X$ and $Y$ within voxels $u$ and $v$, instead of across voxels. This relaxes the spatial coherence constraint while including up to 26 null hypothesis per voxel. Indeed, repeating the experiments of Section~\ref{sec:results} with $k=1$ peaks identifies 330,882 voxels dyads containing 84,147 unique voxels as significant ($p < 10^{-4}$, FDR 0.3\%). These voxels are visualized in Fig.~\ref{fig:within_voxels}, and nearly entirely encompass the previous results in Section~\ref{sec:results} occurring in the same regions but taking larger extent.

\section{Controlling for morphological similarity}
\label{sec:appendix_jacobians}
We seek to identify regions whose similarity between MZ and DZ twins is attributable to oriented white matter microstructure and not simply due to the morphology of the brain or systemic registration misalignment~\citep{kim2008structural}. Brain morphology is known to be heritable~\citep{oppenheim1989magnetic, jansen2015twin}. To that end we identify voxels in which twins have significantly more similar log-jacobian values than strangers do and exclude them from the analysis in this paper. The log-jacobian value of a voxel measures how much this voxel was expanded or contorted from a subject's native space to the normalized space in which the population analysis is performed in.

\begin{figure}[ht]
\centering
\includegraphics[width=0.49\textwidth]{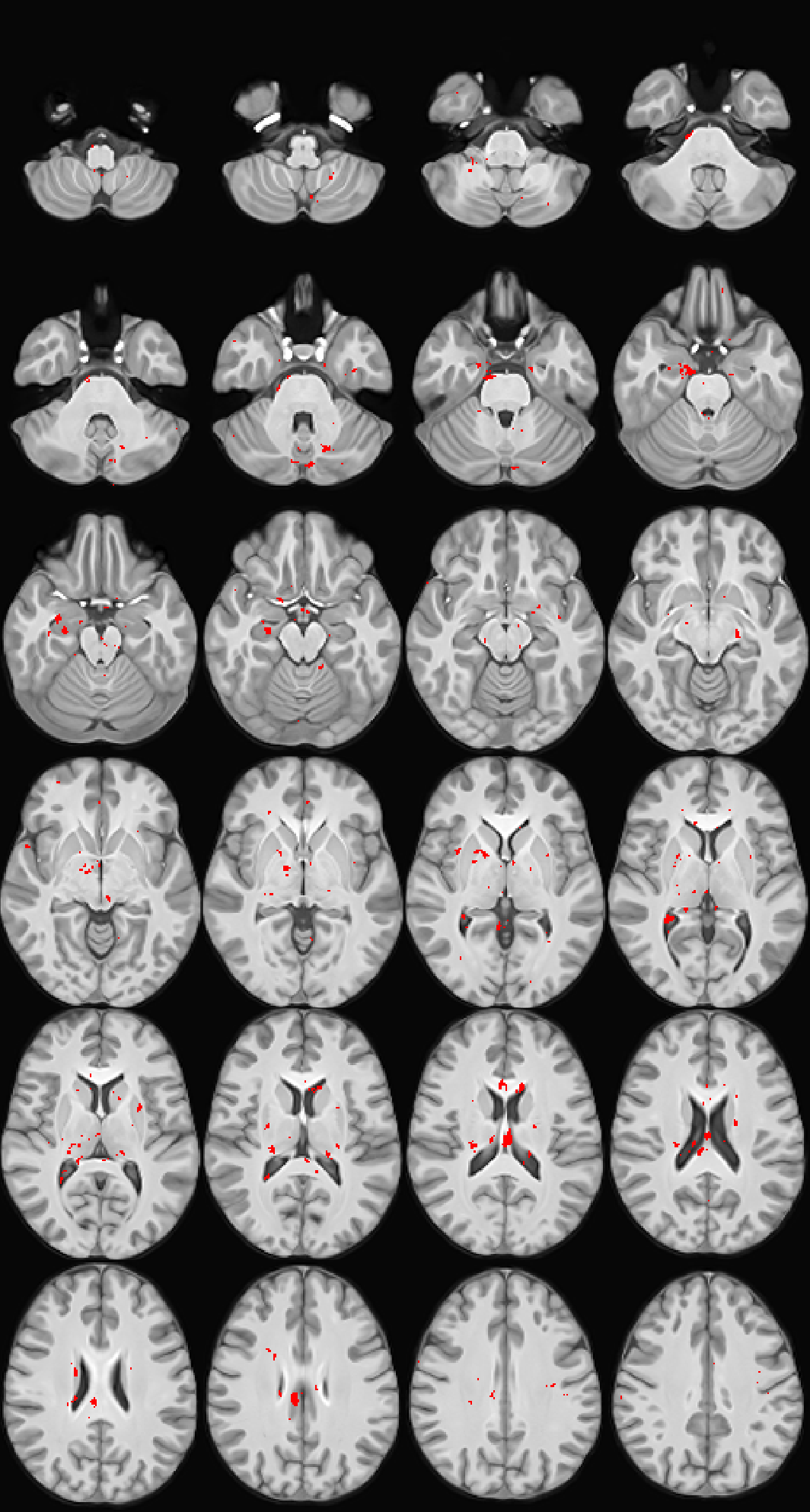}
\caption{Axial slices of voxels in red whose log-jacobian values from the registration process are found to be significantly more similar within monozygotic and dizygotic twins than strangers, suggesting possible morphological similarity. Background image is a population averaged T1 weighted MRI image. Image created partly using ITK-SNAP~\citep{py06nimg}.}
\label{fig:jacobians}
\end{figure}

We define dissimilarity between a pair of subjects with respect to their log-jacobian values as their absolute difference. We compute a distribution of dissimilarities per voxel for twins and strangers and perform a Mann-Whitney U test~\citep{mann1947test} to test if MZ and DZ twins are significantly more similar than strangers in a given voxel. We identified 3.0k voxels that fit this criteria ($p < 10^{-3}$, FDR 28.9\%) using a conservative threshold. Figure~\ref{fig:jacobians} shows an overview of these voxels. Some of the regions identified as such have previously been reported to have heritable anatomical structure~\citep{oppenheim1989magnetic}.

\end{document}